\documentclass[aip,amsmath,amssymb,reprint]{revtex4-1}

\usepackage{dcolumn}
\usepackage{bm}
\usepackage[utf8]{inputenc}
\usepackage[T1]{fontenc}
\usepackage{mathptmx}
\usepackage{lipsum}
\usepackage{xspace}
\usepackage{xcolor}

\newif\ifpdf
\ifx\pdfoutput\undefined
   \pdffalse
\else
   \pdfoutput=1
   \pdftrue
\fi
\ifpdf
   \usepackage{graphicx}
   \usepackage{epstopdf}
\else
   \usepackage{graphicx}
\fi
\graphicspath{ {Figures/} }

\usepackage[english]{babel}

\newcommand{\lao}{LaAlO\textsubscript{3}\xspace}
\newcommand{\sto}{SrTiO\textsubscript{3}\xspace}
\newcommand{\alox}{AlO\textsubscript{x}\xspace}
\newcommand{\insitu}{\textit{in situ}\xspace}

\newcommand{\degreeC}{$^{\circ}$C\xspace}
\newcommand{\etal}{\textit{et al.}\xspace}
\newcommand{\epsr}{\epsilon\textsubscript{r}\xspace}
\newcommand{\vds}{V\textsubscript{\textsc{ds}}\xspace}
\newcommand{\vgs}{V\textsubscript{\textsc{gs}}\xspace}
\newcommand{\cgs}{C\textsubscript{\textsc{gs}}\xspace}
\newcommand{\id}{I\textsubscript{\textsc{d}}\xspace}
\newcommand{\ig}{I\textsubscript{\textsc{g}}\xspace}
\newcommand{\cstray}{C\textsubscript{s}\xspace}
\newcommand{\cchan}{C\textsubscript{c}\xspace}
\newcommand{\micron}{$\mu$m\xspace}
\newcommand{\vth}{V\textsubscript{th}\xspace}
\newcommand{\dlao}{d\textsubscript{\textsc{lao}}\xspace}

\newcommand*{\citen}[1]{%
  \begingroup
    \romannumeral-`\x 
    \setcitestyle{numbers}%
    \cite{#1}%
  \endgroup   
}

\begin{document}


\title{Efficient charge modulation in ultrathin \lao-\sto field-effect transistors}

\author{A.E.M. Smink}
	\affiliation{MESA+ Institute for Nanotechnology, University of Twente, P.O. Box 217, 7500 AE Enschede, The Netherlands}
\author{B. Prabowo}
	\affiliation{MESA+ Institute for Nanotechnology, University of Twente, P.O. Box 217, 7500 AE Enschede, The Netherlands}
\author{B. Stadhouder}
	\affiliation{MESA+ Institute for Nanotechnology, University of Twente, P.O. Box 217, 7500 AE Enschede, The Netherlands}
\author{N. Gauquelin}
	\affiliation{MESA+ Institute for Nanotechnology, University of Twente, P.O. Box 217, 7500 AE Enschede, The Netherlands}
	\affiliation{EMAT, University of Antwerp, Groenenborgerlaan 171, 2020 Antwerp, Belgium}
\author{J. Schmitz}
	\affiliation{MESA+ Institute for Nanotechnology, University of Twente, P.O. Box 217, 7500 AE Enschede, The Netherlands}
\author{H. Hilgenkamp}
	\affiliation{MESA+ Institute for Nanotechnology, University of Twente, P.O. Box 217, 7500 AE Enschede, The Netherlands}
\author{W.G. van der Wiel}
	\affiliation{MESA+ Institute for Nanotechnology, University of Twente, P.O. Box 217, 7500 AE Enschede, The Netherlands}

\date{\today}

\begin{abstract}
At the \lao-\sto interface, electronic phase transitions can be triggered by modulation of the charge carrier density, making this system an excellent prospect for the realization of versatile electronic devices. Here, we report repeatable transistor operation in locally gated \lao-\sto field-effect devices of which the \lao dielectric is only four unit cells thin, the critical thickness for conduction at this interface. This extremely thin dielectric allows a very efficient charge modulation of ${\sim}3.2\times10^{13}$ cm$^{-2}$ within a gate-voltage window of $\pm1$ V, as extracted from capacitance-voltage measurements. These also reveal a large stray capacitance between gate and source, presenting a complication for nanoscale device operation. Despite the small \lao thickness, we observe a negligible gate leakage current, which we ascribe to the extension of the conducting states into the \sto substrate. 
\end{abstract}

\maketitle
\noindent Charge modulation in field-effect transistors (FETs) is the core physical mechanism enabling modern-day electronics. In a standard semiconductor such as silicon, its main purpose is to change the electrical conductivity, defining the ``0'' and ``1'' states in digital electronics. In other classes of materials, e.g., transition metal dichalcogenides and complex oxides, tuning the charge carrier density can trigger quantum phase transitions, offering possibilities for fundamental studies and for using such transitions in electronic devices  \cite{mannhart_high-Tc_1996, ahn_electric_2003, mannhart_oxide_2010}. However, these transitions mostly take place at very high charge carrier densities, exceeding $10^{14}$ cm$^{-2}$. Significant tuning of such a high charge carrier densities can only be achieved by means of chemical doping techniques and electrolyte gating, which are both impractical for functional devices.

In doped strontium titanate (\sto) and at the conducting interface between \sto and selected other insulators such as \lao, (super)conducting and insulating phases are near each other in terms of charge carrier density \cite{schooley_dependence_1965, caviglia_electric_2008, schneider_electrostatically-tuned_2009, liao_metal-insulator_2011}, typically in the range of a few times $10^{13}$  cm$^{-2}$. In the interface case, the geometry is intrinsically the same as the semiconductor-oxide stack of a metal-oxide-semiconductor FET (MOSFET), making such interfaces appealing for use in field-effect devices. Moreover, at low temperatures the \sto substrate can also be used as a gate dielectric (backgating), owing to its huge permittivity \cite{weaver_dielectric_1959}. For the archetypical \lao-\sto interface, reports on the significant field-effect tuning of the critical temperature for superconductivity \cite{caviglia_electric_2008}, the considerable low-temperature mobility \cite{bell_dominant_2009}, and of spin-orbit coupling strength \cite{caviglia_tunable_2010, ben_shalom_tuning_2010} showed the versatility of this system both for fundamental studies and for its possible use in future electronics.

However, backgating takes place over large areas and requires the application of voltages in the order of 100 V across the typically 0.5-mm-thick \sto substrate to achieve a carrier density modulation of up to ${\sim} 4 \times 10^{13}$ cm$^{-2}$ electrostatically \cite{caviglia_electric_2008}. Hence, the backgating geometry is unsuitable for integration into circuits, which requires local operation by voltages of ${\sim}1$ V. Achieving a MOSFET-like (topgate) geometry with the \lao-\sto interface, where the voltage is applied across the \lao layer, is challenging because structuring these materials into (small) channels is not a trivial process \cite{schneider_microlithography_2006}. Low-voltage, topgate FETs were first realized by F\"org and coworkers \cite{forg_field-effect_2012}, after which the functionality of such devices was extended greatly \cite{hosoda_transistor_2013, eerkes_modulation_2013, jany_monolithically_2014}. Unfortunately, the emergence of gate leakage currents across the thin \lao layer -- typically 8 to 20 unit cells (uc) thick -- limited the charge modulation to about $2 \times 10^{13}$ cm$^{-2}$. In parallel, the capability of achieving extreme charge modulations in \sto-based FETs, up to a record value of $2.4 \times 10^{14}$ cm$^{-2}$, was demonstrated in inverted structures with a thick \sto layer as the dielectric \cite{boucherit_extreme_2013, boucherit_modulation_2014, verma_au-gated_2014, verma_large_2016}. Still, these devices have thick dielectrics and a high intrinsic carrier density, compromising low-voltage and local operation. The ultimate oxide-based FET, in which a small gate voltage achieves a charge modulation of several times $10^{13}$ cm$^{-2}$, possibly enabling local switching of quantum phase transitions, therefore has remained elusive.

In this Letter, we demonstrate such efficient charge modulation in Au-\lao-\sto FETs in which the \lao dielectric has a nominal thickness of only four unit cells (uc), the critical thickness for interface conduction \cite{thiel_tunable_2006}. The devices display repeatable transistor behavior with low leakage currents and high ON/OFF ratios. Capacitance-voltage measurements reveal a large voltage-independent contribution to the capacitance, and a low effective permittivity for the \lao layer. The latter can be ascribed to a dielectric `dead layer' forming on the Au-\lao interface, as indicated by scanning transmission electron microscopy. Despite this layer, the charge modulation is very efficient with a high capacitance per unit area, proving the principle of low voltage modulation of high charge densities in complex-oxide based FETs.

The fabrication of our devices started with a standard procedure to terminate the \sto substrate on the TiO$_2$ sites of the (001) surface plane \cite{koster_quasi-ideal_1998}. To enable structuring of the \lao film into channels, we deposited an \alox layer at room temperature, which was etched in OPD4262 developer used for UV lithography \cite{banerjee_direct_2012}. The subsequent growth of \lao and Au by pulsed laser deposition (PLD) was done \insitu, ensuring the interface between these layers to be as clean as possible. The \lao was deposited in an O$_2$ process pressure of $1 \times 10^{-4}$ mbar at $T=850$ \degreeC, with a laser fluence of 1.3 J cm$^{-2}$, spot size of 2 mm$^2$ and a frequency of 1 Hz, resulting in a growth rate of one uc per ${\sim}20$ pulses, monitored by reflective high-energy electron diffraction (RHEED). The distance between the substrate and the single-crystalline target was 45 mm. After the \lao deposition, the sample was annealed for 1 h in an O$_2$ pressure of 400 mbar at a temperature of 600 \degreeC. Then, the Au was deposited in an Ar process pressure of 0.22 mbar, at $T=100$ \degreeC with a laser fluence of 3.6 J cm$^{-2}$ and a spot size of 1 mm$^2$. To reduce the energy of the particles arriving at the substrate, the target-substrate distance was increased to 60 mm. This way, a layer of ${\sim}30$ nm was grown after 9000 pulses at $f = 5$ Hz. Then, electrical contacts to the interface were patterned using UV lithography, followed by a standard technique using Ar ion etching and subsequent sputtering of Ti/Au contacts \cite{eerkes_modulation_2013, richter_interface_2013}, which were structured by lift-off. Finally, the gate electrode was patterned using UV lithography and structured using a buffered KI solution \cite{richter_interface_2013}. 

In the dc current-voltage measurements, the drain current, $\id$, was measured by a Keithley 2401 source-measure unit that also provided the drain-source voltage, $\vds$. The gate current, $\ig$, was determined by measuring the voltage over a $1$ k$\Omega$ resistor using a Keithley 2000 multimeter. The capacitance-voltage characteristics were measured with a Keithley 4200-SCS parameter analyzer with a 4210 capacitance-voltage unit, using an ac voltage of $25$ mV and a frequency of $10$ kHz: close to the optimal frequency of ${\sim}30$ kHz for these devices \cite{schmitz_rf_2003}. All measurements were performed at room temperature with the source terminal connected to ground.

\begin{figure}
\includegraphics{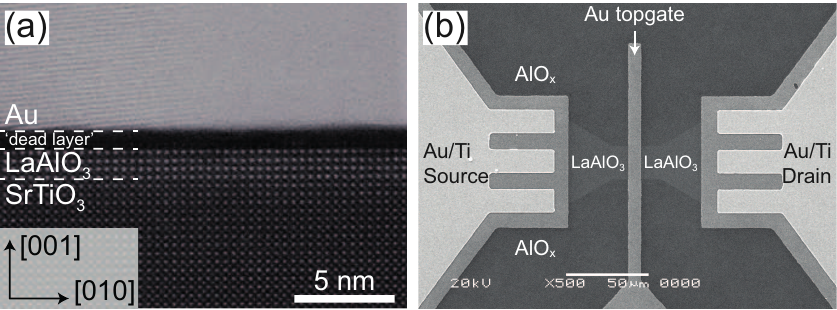}
\caption{(a) High-Angle Annular Dark Field (HAADF) Scanning Transmission Electron Microscopy (STEM) image taken along the [100] direction of a \sto-\lao-Au stack. (b) Scanning Electron Micrograph (SEM) of a FET with width, $W = 20$ \micron and length, $L = 10$ \micron, and indications of the source, drain and (top)gate contacts.}
\label{fig:dev}
\end{figure}

Figure \ref{fig:dev} presents electron microscopy images of two of our devices. In the cross-sectional image (Fig. \ref{fig:dev}(a)), four unit cells of \lao are clearly visible. Like previously reported by another group\cite{richter_interface_2013}, we also observe a thin disordered layer between the \lao and Au layer with a thickness of ${\sim}0.7$ nm. The dark color indicates the absence of heavy elements such as Au, which appears light in this image; we therefore assume that this layer is not conducting and adds to the effective thickness of the dielectric. Further analysis of this layer is to be published elsewhere; we discuss the consequences of this layer for the transistor properties below. Figure \ref{fig:dev}(b) shows the top view of a FET.

\begin{figure}
\includegraphics{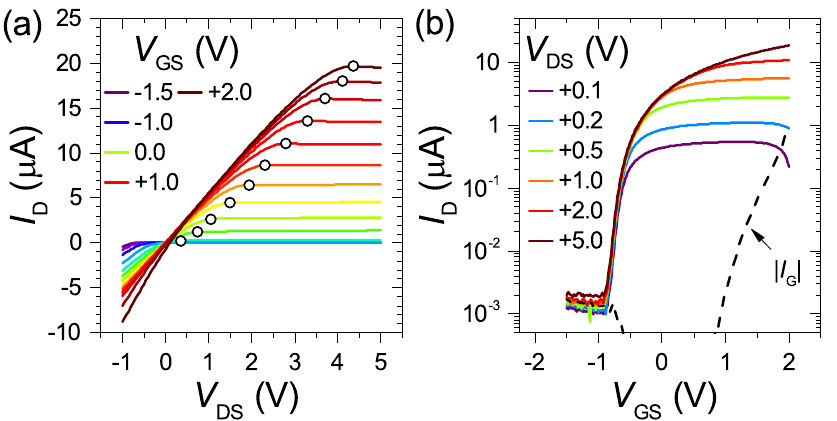}
\caption{Current-voltage characteristics of device A. (a) Drain current, $\id$, versus drain-source voltage, $\vds$, with 250-mV steps in the gate-source voltage, $\vgs$. The open symbols separate the ohmic and saturation regimes for each $\vgs$. (b) Transfer curves for varying $\vds$, and the gate current (dashed line) for $\vds = 0$ V.
}
\label{fig:curvol}
\end{figure}

Figure \ref{fig:curvol} summarizes the transistor operation of device A, a Au-\lao-\sto FET like the one shown in Fig. \ref{fig:dev}(b), with a 4-uc-thin dielectric and channel length, $L$, and width, $W$, both equal to 10 \micron. We measured over ten devices on two different samples, with varying channel dimensions. All of these devices displayed transistor behavior, with ON/OFF ratios between $10^2$ and $10^4$. In Fig. \ref{fig:curvol}(a), we observe clear ohmic (triode/linear) and saturation (active) regimes. The circles separating these regimes represent the saturation voltage and current, which both increase monotonously with $\vgs$. For higher gate voltages, the saturation current does not follow the expected quadratic trend anymore \cite{arora_mosfet_2007}, which we ascribe to a suppression of carrier mobility with increasing topgate voltage as previously observed \cite{hosoda_transistor_2013}. 

In Figure \ref{fig:curvol}(b), switching of the channel conductivity is clearly demonstrated for all $\vds$, from which we extract the transfer characteristics of this device. From a linear fit to $\sqrt{\id}$ versus $\vgs$, we determine the threshold voltage, $\vth$, at $-0.81 \pm 0.01$ V. The subthreshold swing of $98 \pm 2$ mV per decade -- of which the minimum lies at $\vth$, thus is not strictly $sub$threshold -- and the maximum ON/OFF ratio of ${\sim}10^4$ for $\vds = +5$ V are quite comparable to the first semiconductor MOSFETs with a similar dielectric thickness \cite{sasaki_1.5_1996}. Note that the ON/OFF ratio is limited by a finite OFF current, caused by a (minute) gate current emerging below $\vgs = -0.7$ V.

\begin{figure}
\centering
\includegraphics{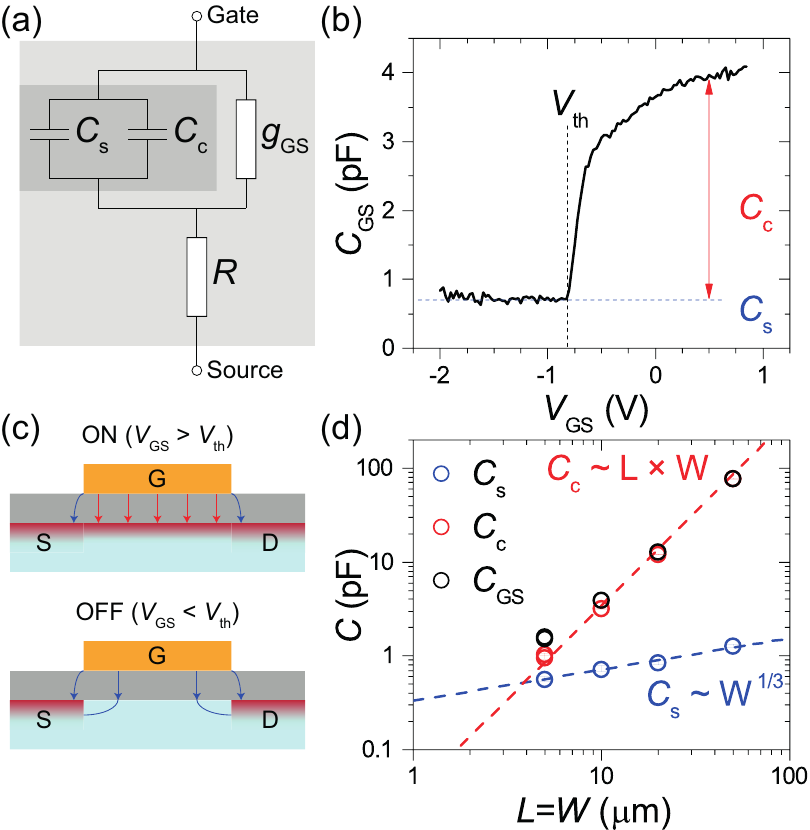}
\caption{Capacitance of Au-\lao-\sto FETs. (a) Equivalent circuit for the gate-source connection of a FET. The gate-source capacitance $\cgs(\vgs)$ is modeled as a voltage-dependent element, $\cchan(\vgs)$ in parallel to a voltage-independent component, $\cstray$. (b) Capacitance-voltage characteristic of device A, with indications of the threshold voltage, $\vth$, $\cchan$, and $\cstray$. (c) Illustration of the electric field lines in the device contributing to the stray (blue) and channel (red) capacitance, above (top panel) and below (bottom) threshold. (d) Scaling of capacitance with channel width, $W$, for devices with $L = W$. Dashed lines are power-law fits to the data for $\cchan$ and $\cstray$.}
\label{fig:cv}
\end{figure}

To further characterize our devices and to determine the charge modulation in the channel, we carried out capacitance-voltage measurements between gate and source. Using a simple model with a shunt and a series resistor next to the capacitance (Fig. \ref{fig:cv}(a)), we extract the capacitance-voltage characteristics of device A. The result, representative for all measured devices, is presented in Figure \ref{fig:cv}(b). The most prominent difference of this $C(V)$ characteristic to semiconductor-based devices is that below threshold, the capacitance remains constant, instead of increasing due to the formation of an accumulation region \cite{arora_mosfet_2007}. This voltage-independent capacitance at negative gate voltage is observed commonly in metal-\lao-\sto junctions \cite{jany_diodes_2010, li_very_2011, singh-bhalla_built-and_2011, keun_kim_capacitancevoltage_2013} and is generally ascribed to a voltage-driven metal-insulator transition (MIT) \cite{jany_diodes_2010}. When comparing this curve to Fig. \ref{fig:curvol}(c), we find that the voltage-independent and voltage-dependent regions are separated by the threshold voltage of $-0.81$ V.

\noindent For a quantitative analysis, we assume that the voltage-independent, or `stray', capacitance, $\cstray$, is a parallel element to the capacitance between the gate and the conducting channel, $\cchan$, yielding $\cgs(\vgs) = \cstray + \cchan(\vgs)$. To substantiate this assumption, we measured the capacitance-voltage characteristics of several devices with varying dimensions. This allows to extract the scaling of gate-source capacitance with device area, as depicted in Figure \ref{fig:cv}(d). Here, we consider devices with a square channel, i.e. $L = W$, and extract the capacitances $\cgs$ and $\cchan$ at $\vgs = +0.5$ V, where the capacitance is fairly constant as function of gate voltage. We find that the total capacitance does not follow a power law dependence on $W$, but that $\cstray$ and $\cchan$ do. The channel capacitance scales with device area ($L \times W$), in excellent agreement with a parallel-plate capacitor model. Hence, we can extract the effective relative permittivity, $\epsr \approx 6.1 \pm 0.4$, using $d = 1.5$ nm. The stray capacitance scales with the channel dimension to the power $1/3$, thus depends on the geometry in a nontrivial way. To our knowledge, there is no theory yet that explains such a dependence on the device geometry. Extrapolation of our data suggests a crossover to occur at $L = W \approx 3.5$ \micron, implying that the stray capacitance becomes dominant in small devices.

We postulate that this large stray capacitance is due to the very large permittivity of the channel material, \sto, which is ${\sim}300$ at room temperature \cite{weaver_dielectric_1959}. As illustrated by the blue lines in Fig. \ref{fig:cv}(c), the gate terminal is capacitively coupled to the source and the drain through an electric field. The capacitance associated with this electric field depends on the permittivity of the insulator and of the channel material. In most materials, this would not be very significant; here, the very high permittivity of the channel material implies that the gate-source capacitance in absence of a conducting channel remains sizable.

To obtain the modulation of charge density in the channel, we omit the stray capacitance and integrate $\cchan$ with respect to $\vgs$. Between threshold and $+1$ V, this yields a carrier density modulation of $3.2\times10^{13}$ cm$^{-2}$. Above $+1$ V, the measurement becomes inaccurate due to emerging gate leakage. If we assume that the capacitance remains constant above $+1$ V, the projected charge modulation during the measurements presented in Fig. \ref{fig:curvol}(b) is ${\sim}5.2\times10^{13}$ cm$^{-2}$. The gate voltage required for this modulation does not exceed $+2$ V; within this window, the gate leakage current remains more than an order of magnitude smaller than the current through the channel.


As a benchmark to compare these FETs to silicon-based devices, we calculate the equivalent oxide thickness (EOT) compared to SiO$_2$, which has $\epsr = 3.9$. Using $\epsr \approx 6.1 \pm 0.4$ and $d = 1.5$ nm, we find an EOT of only $0.96 \pm 0.1$ nm. However, this value for $\epsr$ is much lower than ones reported in literature for thick \lao films \cite{edge_electrical_2006, robertson_high_2004}, which range from $18$ to $30$. This suppression of $\epsr$ in metal-\lao-\sto junctions is a widely observed phenomenon \cite{jany_diodes_2010, li_very_2011, singh-bhalla_built-and_2011, hosoda_transistor_2013, keun_kim_capacitancevoltage_2013}; with increasing \lao layer thickness, $\epsr$ was reported to approach the bulk value \cite{hosoda_transistor_2013}. This was ascribed to a `dead layer' forming inside the \lao film because of structural interface effects \cite{stengel_origin_2006}. Because Fig. \ref{fig:dev}(a) shows a disordered layer forming on top of the fully intact, 4-uc-thin \lao layer, we propose an alternative scenario in which the `dead layer' forms on top of the \lao rather than inside it. A series capacitor model in this scenario is mathematically equivalent to the one used by Hosoda \etal \cite{hosoda_transistor_2013}:

\begin{equation}
\frac{\dlao + d\textsubscript{dead}}{\epsr\textsubscript{,tot}} =  \frac{\dlao}{\epsr\textsubscript{\textsc{,lao}}} + \frac{d\textsubscript{dead}}{\epsr\textsubscript{,dead}}.
\end{equation}

\noindent By using $\epsr\textsubscript{\textsc{,lao}} = 18$, $\dlao = 1.52$ nm, and $d\textsubscript{dead} = 0.7$ nm, we find $\epsr\textsubscript{,dead} \approx 2.5 \pm 0.3$, which we deem a reasonable value for a disordered layer. Therefore, this disordered layer at the \lao-Au interface poses a viable alternative scenario to the dead layer within the \lao film; further investigations on devices with varying \lao layer thickness made using different fabrication procedures may distinguish between these two possibilities.

\begin{figure}
\includegraphics{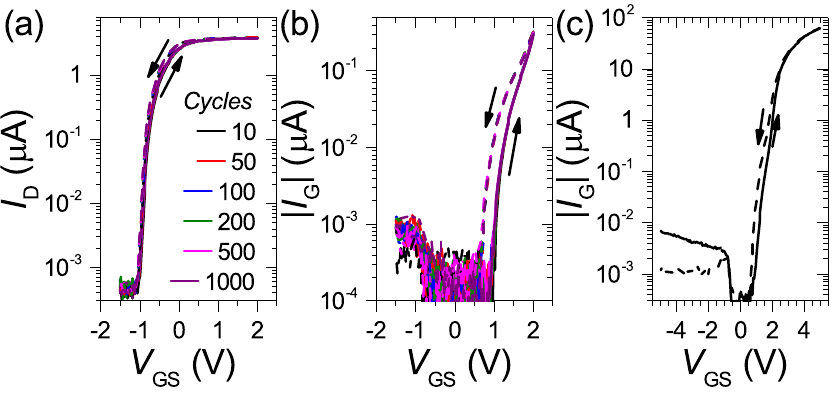}
\caption{(a) Response of the transfer curve of device B ($L = W = 5$ \micron) to repeated gate voltage cycling between $\vgs = -1.5$ V and $+2$ V, with $\vds = +1$ V. Solid (dashed) lines represent sweeping $\vgs$ upwards (downwards). (b) Gate current of device B during the repeated cycling in (a), for $\vds = 0$ V. (c) Gate current of device C ($L = W = 10$ \micron) upon sweeping the gate voltage to $\pm5$ V, for $\vds = 0$ V.}
\label{fig:stab}
\end{figure}

Despite the suppression of $\epsr$, the charge modulation in our devices is efficient and not compromised by gate leakage currents. Moreover, in comparison to previous reports on Au-\lao-\sto stacks with 4-uc-thin barriers, where the gate current was used to perform tunneling spectroscopy of the interface \cite{richter_interface_2013, boschker_electronphonon_2015, kurten_-gap_2017}, the gate current density is about four orders of magnitude smaller. To investigate the factors determining the gate leakage current, we carried out temperature-dependent measurements as described in the Supplementary Information. The results show that ohmic and hopping-based conduction are negligible, and that the gate current is dominated by direct tunneling and Schottky emission. Accordingly, the density of defects inside the \lao layer must be very low and memristive effects based on the movement of defects should be absent. To confirm this, we measured the device response against repeated gate voltage cycling and up to high gate voltages of $\pm 5$ V. The results presented in Fig. \ref{fig:stab} confirm the absence of resistive switching in our devices. Moreover, the behavior at high voltage in Fig. \ref{fig:stab}(c) fits the description of a Schottky diode with a forward-reverse bias ratio of ${\sim}3 \times 10^4$, in good agreement with previous results \cite{jany_diodes_2010}.

To explain the surprisingly large difference in gate current between the tunneling spectroscopy devices of Refs. \citen{richter_interface_2013, boschker_electronphonon_2015, kurten_-gap_2017} and our FETs, we consider the factors determining direct tunneling currents. Since both types of devices have the same material stack with the same thicknesses and the `dead layer' is present in the tunneling spectroscopy studies \cite{richter_interface_2013} as well, the energy landscape in terms of barriers and thickness should be the same. Therefore, we argue that the out-of-plane distribution of mobile charges in the \sto differs greatly between the two types of devices, increasing or decreasing the effective barrier thickness. In \sto, the mobile charges do not reside exactly at the surface, but are distributed within a quantum well \cite{biscaras_two-dimensional_2012, gariglio_electron_2015, smink_gate-tunable_2017}. This charge distribution depends crucially on the electrostatic boundary conditions for the well, which are highly susceptible to the environment in which the \lao film is grown \cite{gunkel_space_2016}. In consequence, samples grown in different conditions have varying charge distributions in the quantum well. Especially for samples with thin \lao layers, the effective depth at which the mobile charges reside can thus vary greatly among samples grown under different conditions. We note that this effect may be enhanced greatly by a recently proposed region of negative polarization directly on the \sto side of the interface \cite{raslan_possible_2018}. Of great importance for device operation is that this effective thickness increase should not suppress the capacitance by much, for the permittivity of \sto exceeds 300 even at room temperature \cite{weaver_dielectric_1959}. Hence, we propose that this increase of the effective tunnel barrier thickness, without lowering the capacitance, is the key enabler of the efficient charge modulation observed in our devices.


In summary, we characterized the operation of Au-\lao-\sto field-effect transistors with a \lao layer thickness of only four unit cells, or 1.5 nm. Our devices exhibit highly repeatable transistor behavior with very low gate leakage currents. In capacitance-voltage measurements, the gate-source capacitance becomes voltage-independent below threshold, which we attribute to stray fields coupling the gate to the source and drain terminals in absence of a conducting channel. Integration of the voltage-dependent part of the capacitance yields a charge modulation of about $3.2\times 10^{13}$ cm$^{-2}$, within a gate voltage range of $\pm 1$ V.

This highly efficient charge modulation is limited by a commonly observed suppression of the permittivity in very thin \lao layers grown on \sto. Scanning transmission electron microscopy imaging suggests that this suppression is due to a dielectric `dead layer' forming at the Au-\lao interface, with a thickness of ${\sim}0.7$ nm. The surprisingly low leakage current cannot be due to this layer, but is likely due to the out-of-plane distribution of charges in the \sto channel. Because of the high dielectric permittivity of \sto, this does not significantly affect the gate-source capacitance, enabling efficient modulation of high charge densities by low gate voltages without excessive gate leakage currents. We foresee that making use of this delocalization in quantum wells opens new venues to engineer high-charge-density field-effect transistors based on advanced materials.

\begin{acknowledgments}
\noindent We thank Maurits de Jong for his help with the capacitance-voltage measurements, Jochen Mannhart and Hans Boschker for stimulating discussions, and Frank Roesthuis, Dick Veldhuis, and Thijs Bolhuis for technical assistance. We acknowledge financial support through the DESCO program of the Foundation for Fundamental Research on Matter (FOM), associated with the Netherlands Organization for Scientific Research (NWO).
\end{acknowledgments}

\bibliography{FETpaper}

\end{document}